\documentclass[a4paper]{jpconf}
\usepackage{graphicx}
\usepackage{citesort}
\usepackage{color}

\usepackage{epstopdf}

\begin{document}
\title{Open issues in  probing 
interiors of solar-like oscillating  main sequence stars:\\
 2. Diversity in the  HR diagram}

\author{MJ Goupil${^1}$,  Y. Lebreton${^1}$, J.P. Marques${^1}$, S.
Deheuvels${^2}$, O. Benomar$^3$, J.
Provost$^4$}

\address{$^1$ Observatoire de Paris,  UMR 8109, Paris, France}
\address{$^2$ Astronomy Department, Yale University, New Haven CT, U.S.A.}
\address{$^2$ Sydney Institute for Astronomy (SIfA), School of Physics, University of 
Sydney, NSW 2006, Australia}
\address{$^3$ Universit\'e de Nice-Sophia Antipolis, CNRS UMR 6202, Observatoire de la Côte d'Azur,  Nice, France
}

\ead{mariejo.goupil]@obspm.fr}

\begin{abstract}
We review some major open issues in the current modelling of low and intermediate mass, main sequence stars based on seismological studies. 
The solar case was discussed in a companion paper, here 
  several issues specific to other stars than the Sun are 
 illustrated with a few stars observed with CoRoT and expectations from Kepler data.
\end{abstract}

\section{Introduction}
After   more than two  decades  of helioseismology, 
almost four  years  of asteroseismology with CoRoT \cite{baglin06} and almost two years  
 of  intensive asteroseismology  with KEPLER \cite{borucki07}, we review some current open
issues about the internal structure of  solar-like oscillating  stars.
 We  focus on   low and intermediate mass, 
main sequence stars.  We started  with the Sun  and stars that have a similar structure than the Sun in a
companion paper. Here open issues not encountered with the Sun will be discussed 
and illustrated  with a few individual stars  
observed by CoRoT   and from ground. We end with a brief  discussion  about  expectations from   KEPLER data.
For sake  of  shortness, we decided  to include  only unpublished figures and to 
cite published figures in the text.
Several reviews exist on the topic, for instance \cite{jcd09}, \cite{jcdhoudek10}.

\section{ From the Sun to solar-like oscillating MS stars:}
We focus on low and intermediate mass, main sequence stars that-is  stars with masses  up to 1.5 $M_\odot$ 
 corresponding to  F, G, K spectral types. 
These stars  have an external convective region and  can oscillate like the Sun 
 with high frequency p modes. 
For these stars, one encounters the same problems as for the Sun, namely surface effects  when comparing absolute
values of the frequencies.  But although we refer to these stars as solar-like stars in the present framework  for shortness,
these stars can  differ from the Sun by several aspects: mass, age, surface metallicity   and  helium abundances, 
initial conditions for the chemical 
abundances, the rotation and  magnetic properties.  For masses larger than about 1.2 $M_\odot$, 
they have a convective core  unlike the Sun. 
 These differences therefore  lead to  additional open questions about their modelling. The
major problem concerns dynamical processes occuring inside stars that
have a significant impact on the mean structure of the stars and their ages.
 In particular, the 3D multiscaled turbulent convective transport and
related instabilities is taken into account by means of  a 1D crude
formulation that involves free parameters which cannot be derived from first principles. As a consequence
they are calibrated with observations (the Sun, binaries) but these values have no predictive quality.  

The input parameters  such as  mass, age, initial chemical composition $Y_0, (Z/X)_0$, rotation profile
  are  usually  not well known. A first order of magnitude for  mass and age 
  is obtained   via the  location  of the star in the HR diagram based on  photometric and spectroscopic information.
Uncertainties on this location and the chemical abundances give rise  to a 
large number of possible models which makes difficult  to probe in detail 
the internal structure of the star. This number is significantly 
 reduced when seismic diagnostics such as the large separation and the small spacings are used. 
However free parameters used to describe the convective transport and related instabilities 
   increase the family of acceptable models. The resulting solution 
  therefore remains input physics dependent.
                      
\subsection{Observational constraints and seismic diagnostics}
\begin{figure}[t]
\centering
\resizebox*{0.5\hsize}{!}{\includegraphics*{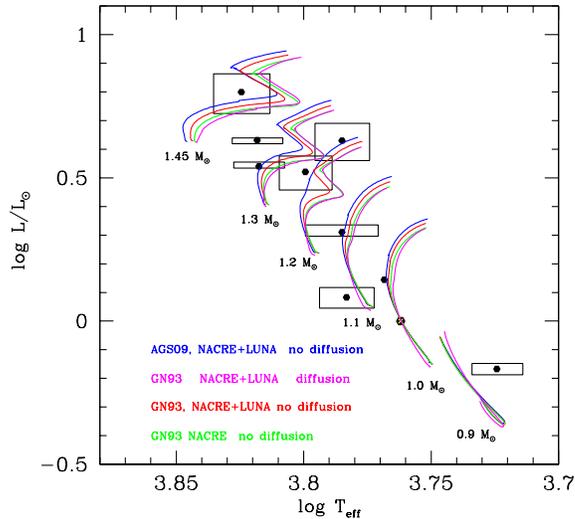}}
\caption{\small     Evolutionary tracks  of models built with several assumptions as indicated
 in the plot. All nuclear reaction rates are from NACRE except for the 
 reaction $^{14}N(p,\gamma)^{15}$  of the CNO cycle  which is from LUNA when specified. 
  The other nuclear rates are all from NACRE. The helium and metallicity are solar. }
\label{figstarmixture}       
\end{figure} 

Efforts are therefore currently put at obtaining seismic constraints  
that are  discriminating and  as model independent  as possible.  
A first information that is looked for is the large separation, 
its mean value as well as 
its variation with frequency
\cite{Mosser09}, \cite{Roxburgh09}.
Other combinations of  low degree mode frequencies are built to locate  the 
base of the upper convective zone   and obtain properties about  He ionization regions
\cite{Monteiro00}; \cite{Mazumdar01a}; \cite{Mazumdar01c};
 \cite{Mazumdar10}  and surface  helium content \cite{Basu04}.      
 Theoretical developements are carried also out
  to devise diagnostics and appropriate methods to  determine the age 
  (\cite{Houdek09} and references therein); 
  to probe the core \cite{Provost93}, \cite{Roxburgh99},\cite{Roxburgh02}; \cite{Roxburgh04};
    \cite{Mazumdar01c}, \cite{Mazumdar06} with specific attention to  tiny convective core properties  
    \cite{Cunha10} of low mass stars  using only    low degree modes; to investigate  mixing 
    beyond the  convective core  \cite{Popielski05}
or semiconvection \cite{SilvaAguirre10}.
For instance, the spacing $d_{01}$ is sensitive to convective core properties 
 \cite{Provost05},  
  \cite{Miglio05}; ; \cite{Roxburgh07}; \cite{Deheuvels10a}. 
   \cite{Deheuvels10a}   shows  the variations of the 
seismic quantity $d_{01}$ with frequency over a large range of frequency for a model representing  the star  HD 203608.
The slope of the observed $d_{01}$ over the observed frequency range is  directly related to the slow period of oscillation seen 
on the theoretical $d_{01}$  that 
corresponds to a time scale  related to the acoustic radius of core convective radius.

\subsection{CoRoT solar-like stars}
Most  CoRoT solar-like stars  
differ from the Sun either  because they are   more massive and faster rotators
such as the F2 star HD181420 \cite{Barban09}; \cite{benomar10phd} and the F5 star  HD170987 
\cite{Mathur10}  or
 because they are   evolved  and have an  isothermal core such as the G0 star HD49385
   \cite{Mosser08},\cite{Deheuvels10b}
or they have a mass similar to that of the Sun  but  differ in their  metallicity 
such as HD49933   \cite{benomar10}, 
 HD 52265,  or the young magnetic star 
 HD 46375  \cite{Gaulme10}. 
 These stars can also  differ significantly from the Sun by 
  their level of magnetic activity and their magnetic cycles \cite{Garcia10}.


\subsubsection{Which star for which diagnostic?}

Due to their differences, some of these 
  stars are  better  suited  to probe different physical processes.\\
{\it Initial abundances and chemical mixture:} Fig.\ref{figstarmixture}  shows some of these  stars in a  HR diagram together with evolutionary tracks 
of models built assuming no microsocopic diffusion; nuclear reaction rates are
from NACRE completed with recent  LUNA  determinations. Two mixtures have been
used : AGS09  and GN93. The effect of changing the
mixture  has a clear impact on a star such that the F star  HD181906  \cite{Garcia09} 
which has a convective core
in the first case and no convective core in the second case. 
As the existence of a convective core manifests itself with a large slope of  $d_{01}$,
this can be in  principle determined with this seismic diagnostic. \\
{\it Nuclear reaction  rate:CNO cycle and  $^{14}N(p,\gamma)^{15}$ burning:}
The change from NACRE \cite{angulo99} to LUNA \cite{Formicola04} reaction rate led to  
reduce the CNO cycle efficiency. 
As a result, the tracks for the most massive stars (i.e.  with central temperature high enough
 for CNO to dominate)   are  slightly shifted upward. This of course coincides with 
 stars having a convective core.  
 For a $1.2 M_\odot$ and  $Z=0.01$ model, the  convective core
is smaller at given mass  and appears at higher mass \cite{angulo99}.\\
{\it Microscopic diffusion, surface helium abundance and initial metallicity:} Due to gravitational  settling and atomic diffusion in the radiative region below the convective envelope, 
the surface helium  decreases with time. This decrease is larger for higher mass 
stars  because the convective envelope is thiner.
 The decrease is also larger for lower metallic stars which are more compact
and - due to smaller opacities and
therefore smoother radiative temperature gradient-  have also a
thiner convective envelope.  The disparition of helium in such  thin
convective zones is therefore  very rapid.  If  microscopic diffusion  acts alone, the
envelope of  HD49933 for instance  would be fully depleted of helium.  Mechanisms opposite to diffusion must
therefore be at work such as 
 turbulent diffusion and/or radiative acceleration, rotationally induced mixing and  must therefore
be included in the modelling.

\subsubsection{Mode degree identification}

Seismic inferences assume that the modes are identified 
that- is that the observed frequencies can be
attributed to  modes with given degree $l$ and azimuthal number $m$ values 
in a spherical harmonics description.  
However in some cases,  
 the  $l=1$  split multiplets
  can be mistaken with  overlapping $l=0$ and
 $l=2$  modes in the fitting process. Hence  some ambiguity  can exist in the determination of 
 $l=0$ and $l=1$ ridges in an echelle diagram. 
This is particularly true for the
hottest  (F type) stars   
because of their large mode linewidhs and their relatively
 fast rotation.
Exemples are  HD49933 \cite{Appourchaux08}, HD181420 
\cite{Barban09}  and HD181906 \cite{Garcia09}  for which 
2 scenarii  are proposed for the identification of the $l=0 $ and $l=1$ modes. 
In one case,  fitting the data imposes   a quite   large core overshoot  
 whereas in the other case
a usual  intermediate core overshoot  amount is sufficient.
This ambiguity that exists when the data sets are too short 
can be  lifted when using   sophisticated data analysis treaments 
\cite{Benomar09a}, \cite{Benomar09b}, {\cite{Gaulme09},  \cite{Mosser09}.
 To lift the ambiguity in mode identification when it exists, \cite{Bedding10}  have proposed 
to use scaling relations ($\nu$ and $\Delta \nu$ scale  
 as $<\Delta \nu>$)    to build   scaled echelle diagrams with reference to  a star similar to the  studied one.
The authors tested this procedure with two sets of twin stars Sun and 18 Sco   -
  $\tau$ Ceti and $\alpha$ Cen B   and used it to  determine the  most probable scenario for 
two CoRoT stars  HD181420  and  HD181906      
using   HD49933    as the reference star. 
 This scaling procedure has then  been recently used on  
 the ground based observed star,  HD203608 \cite{deheuvels10phd}. 
 Indeed for this   F8V   star,
two possible scenarii have been found 
 with again important consequences on conclusions that can be drawn 
from the mode identification \cite{Mosser08};  \cite{Deheuvels10a}. 
Echelle diagrams  built assuming the   scaling according to
  \cite{Bedding10}
   coincide with  the observed echelle diagram
   in the case of one of the two possible scenarii and definitely 
 rejects the alternative scenario \cite{deheuvels10phd}. 
 This scenario favours a  mild overshoot and  the 
 survical of convective core despite the small mass and  old age  of the star
  due its low metallicity \cite{Deheuvels10b}.  Mode identification  based on scaling relations  therefore appears as a potential interesting method 
that must nevertheless be studied further  on theoretical ground   before it can be used 
as a proper decision method.

\begin{figure}[t]
\resizebox*{0.5\hsize}{!}{\includegraphics*{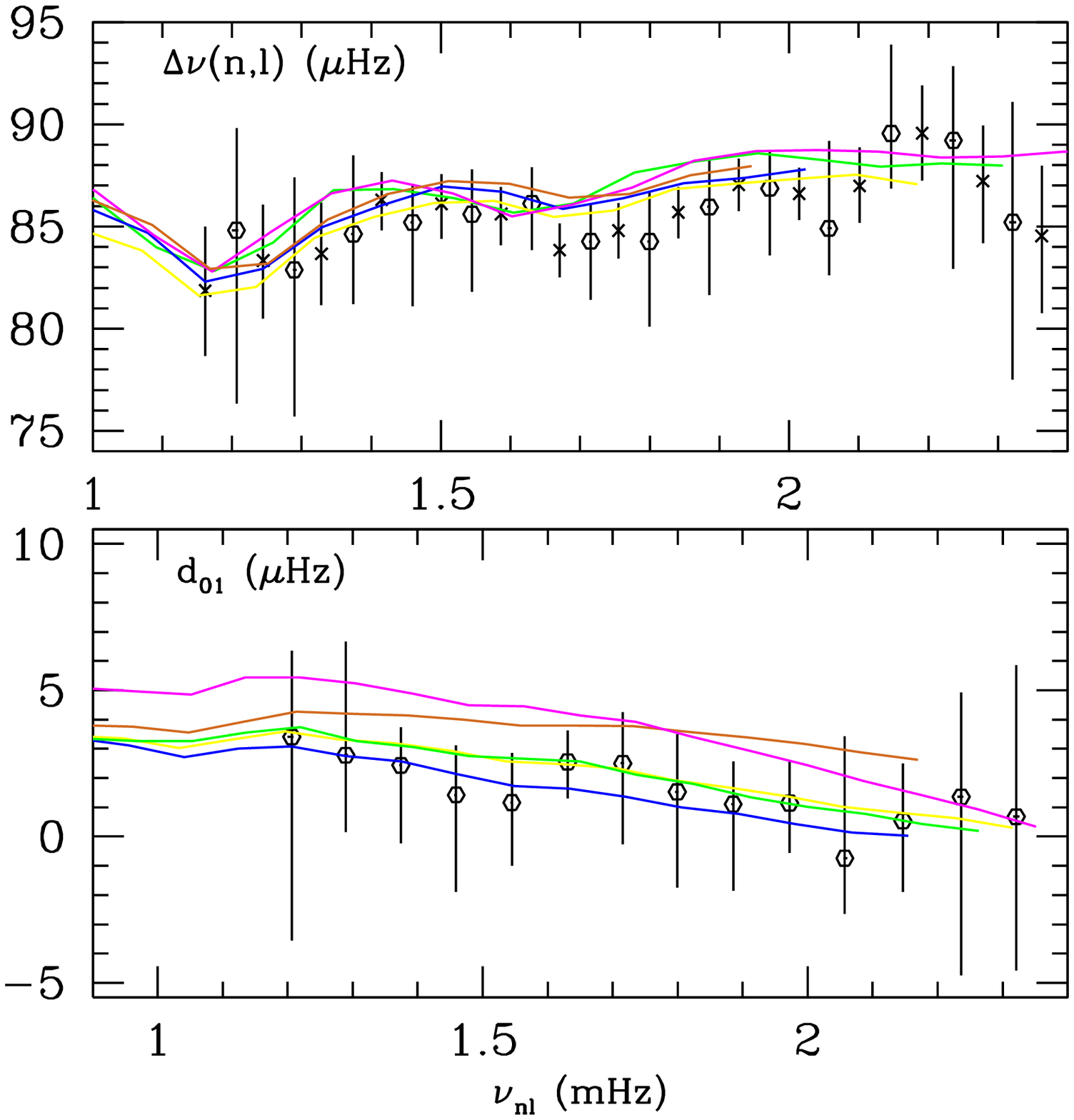}} 
\resizebox*{0.5\hsize}{!}{\includegraphics*{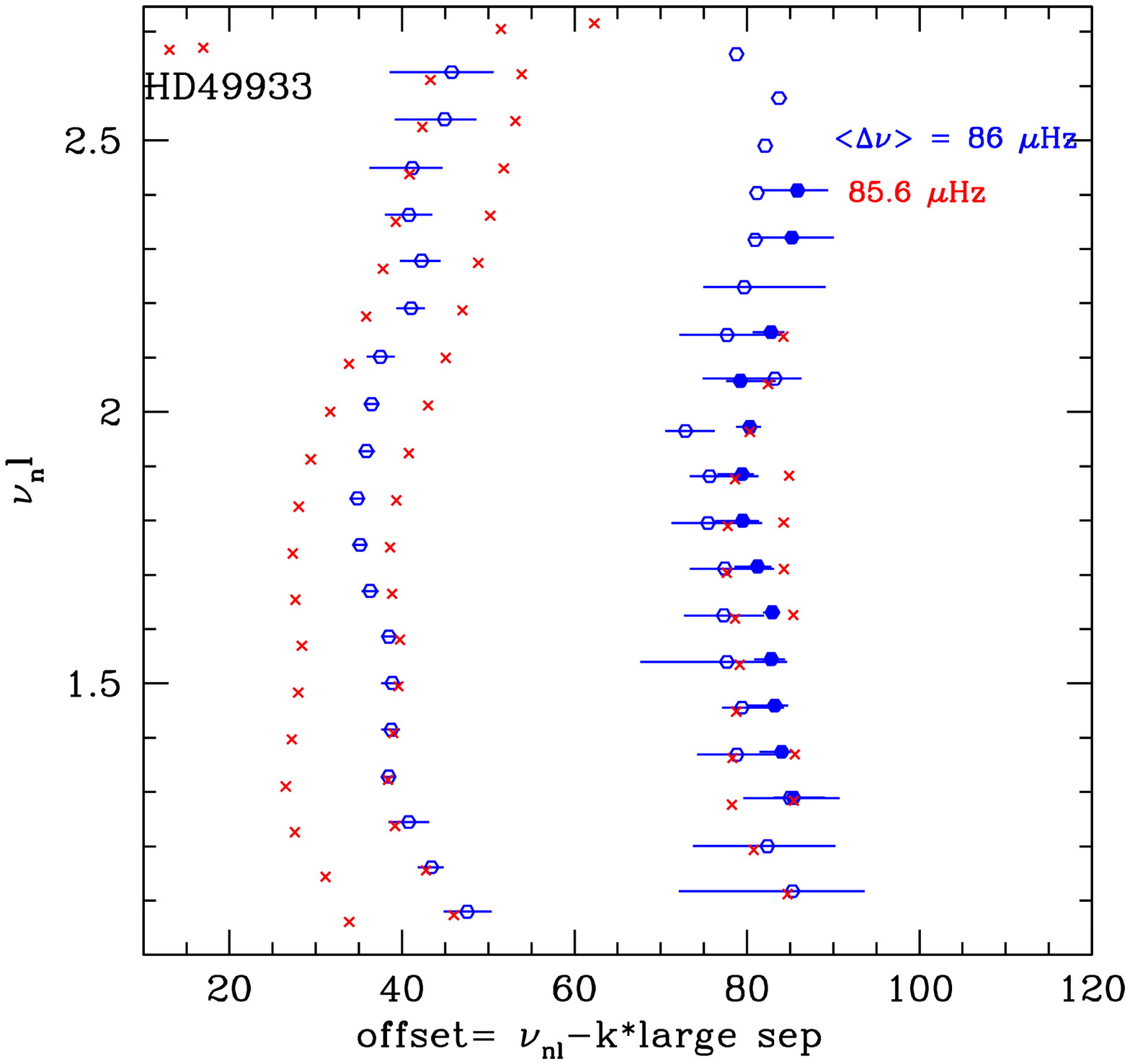}}
\caption{\small {\bf (Left top:)} Large separation $\Delta \nu_{n,l} $ 
in function of the frequency $\nu_{n,l}$ 
and  {\bf (left bottom:)} the small
spacings   $d_{01} =  \nu_{n,0} - (\nu_{n,1}+\nu_{n-1,1})/2$
and $d_{10} = - \nu_{n,1} + (\nu_{n+1,0}+\nu_{n,0})/2$
 in function of the frequency $\nu_{n,0}$  for HD49933.  Data (open circles and crosses) are from \cite{Benomar09b}. 
Models  (solid curves)    are built assuming 
AGS05, no diffusion;  $\alpha_{cgm}=0.6$, $\alpha_{ov}=0.27$ (magenta);
AGS05,  diffusion, $\alpha_{cgm}=0.6$,  $\alpha_{ov}=0.21$,  $Y=0.27$,  $ Z/X= 0.0079$ (yellow) 
AGS09,  diffusion (green);
AGS05,  diffusion,$\alpha_{ov}=0.2$, rotationally induced transport (blue);
AGS05,   diffusion,  $\alpha_{ov}=0.$, rotationally induced transport,
  $Y=0.27$,  $\alpha_{cgm}=0.6$   (chocolate).({\bf Right:})Echelle diagram for HD49933.  Blue  dots represent the  observations
 and  red ones the model.
 }
\label{HD49}       
\end{figure}

   \subsubsection{HD49933: a  low metallicity  star }
   With 180 days of observation   with CoRoT, all seismic data analyses 
agree  to provide the same   scenario   for the $l=0,1$ 
 mode identification  for this star \cite{benomar10}. 
 Seismic modelling of HD49933  illustrates   the difficulty one encounters 
because of  the degeneracy in input  parameter space
composed of   mass, age, $(Z/X)_0,Y_0$, $\alpha_{cgm}, \alpha_{ov}$, initial rotation  and transport
  coefficients. The free parameters $\alpha_{cgm}, \alpha_{ov}$ represent the convective 
  mixing length for the adopted CGM formulation \cite{cgm96} and the convective core overshoot parameter respectively.
    All models  discussed below are calibrated so that 
the mean large separation  $<\Delta \nu>$ and  the mean 
  small spacings $d_{01},d_{10}$   agree with the observations as well as the observed  location in a HR
  diagram. 
 This lifts only partially the degeneracy. 
    The   phase of oscillation of the large separation is   found to 
be    quite  sensitive to values of $\alpha_{cgm}$ and  $Y_0$
and therefore such a constraint must be added in the minimisation process to reduce the number of
acceptable models although this has not been done  here. 
Individual frequencies have not been fitted (see below).
The surface metallicity of the star is  taken  in the  observed range 
  $[Fe/H]=-0.4 \pm 0.1 $, significantly  lower than the Sun \cite{Bruntt09}.  

Fig.\ref{HD49} compares the mean large separation and the small spacings $d_{01}, d_{1,0}$  for models built
assuming either  AGS05 or AGS09 mixtures and  various assumptions about microscopic diffusion,
 rotationally induced transport, convective core
overshoot   as listed in the caption. All models are computed with CESAM2k
 \cite{morel08}. Models including   rotationally induced  transport of angular momentum
as  implemented by J. Marques  in
 the code CESTAM  (a modified version of cesam2k)  have been computed assuming no loss of
 angular momentum. The initial angular rotation on the PMS  has been set  
in order  to fit the observed rotation period of the star, 
$P=3.4$ days at the age of HD49933. Details on  the modelling of this star will be published  elsewhere. 

 For all these different assumptions, one can find a  mass and age that fit the mean  large separation 
oscillation by ajusting $\alpha_{cgm}$ and $Y_0$  although this has 
not  been done yet for our rotating models. 
The mass is  found in the range 1.05-1.18 $M_\odot$ and the age in the range 
2900-3900 Myr depending  on the
  assumptions in the physical description and the chemical abundances.
As a result of these various calculations, we find that 
 a) when  the AGS05  mixture is assumed, 
it is  difficult to find a model satisfying all the observational constraints 
when $(Z/X)_0$ is on the smaller part of the authorized interval. This is less the case with the 
 less extreme AGS09 mixture. It is important to stress that  the star
  being metallic  deficient compared with the Sun,
 the above  mixtures taken from the Sun	might not well be suited;
 b) because this star is low metallic,
 its thin  convective envelope is rapidly devoid of helium 
when  microscopic diffusion is included   if one starts with solar initial  helium 
abundance
when one  assumes AGS05 mixture.
One then  needs to start with a  large initial helium
abundance $Y_0$ 
  or  one must  include   some turbulence in the radiative zone.
 Starting then with a large $Y_0=0.35$, one  still  obtains a  small  $Y_{surf}=0.10$  
 value   for HD49933.
One obtains a  less extreme surface helium abundance  $Y_s=0.18$
when using  the less extreme AGS09 mixture; 
c) without including microscopic diffusion
nor rotationally induced transport, no model fits $d_{01}$ for an overshoot smaller that $0.25-0.3 ~Hp$
whatever the mixture;
 d) for AGS05 mixture, when diffusion is included with or without 
  rotationally induced transport, some intermediate amount of overshoot ($\approx 0.2 ~Hp$) 
  still remains  necessary.  
As a conclusion,   microscopic  diffusion
 and  rotationally
induced transport as modelled here are not enough to render count of the slope
of $d_{01}$ variation with AGS05. One needs to include also some amount 
of convective core overshoot   as a proxy 
for  true overshoot and/or   additional mixing process.

\begin{figure}[t]
\resizebox*{0.5\hsize}{!}{\includegraphics*{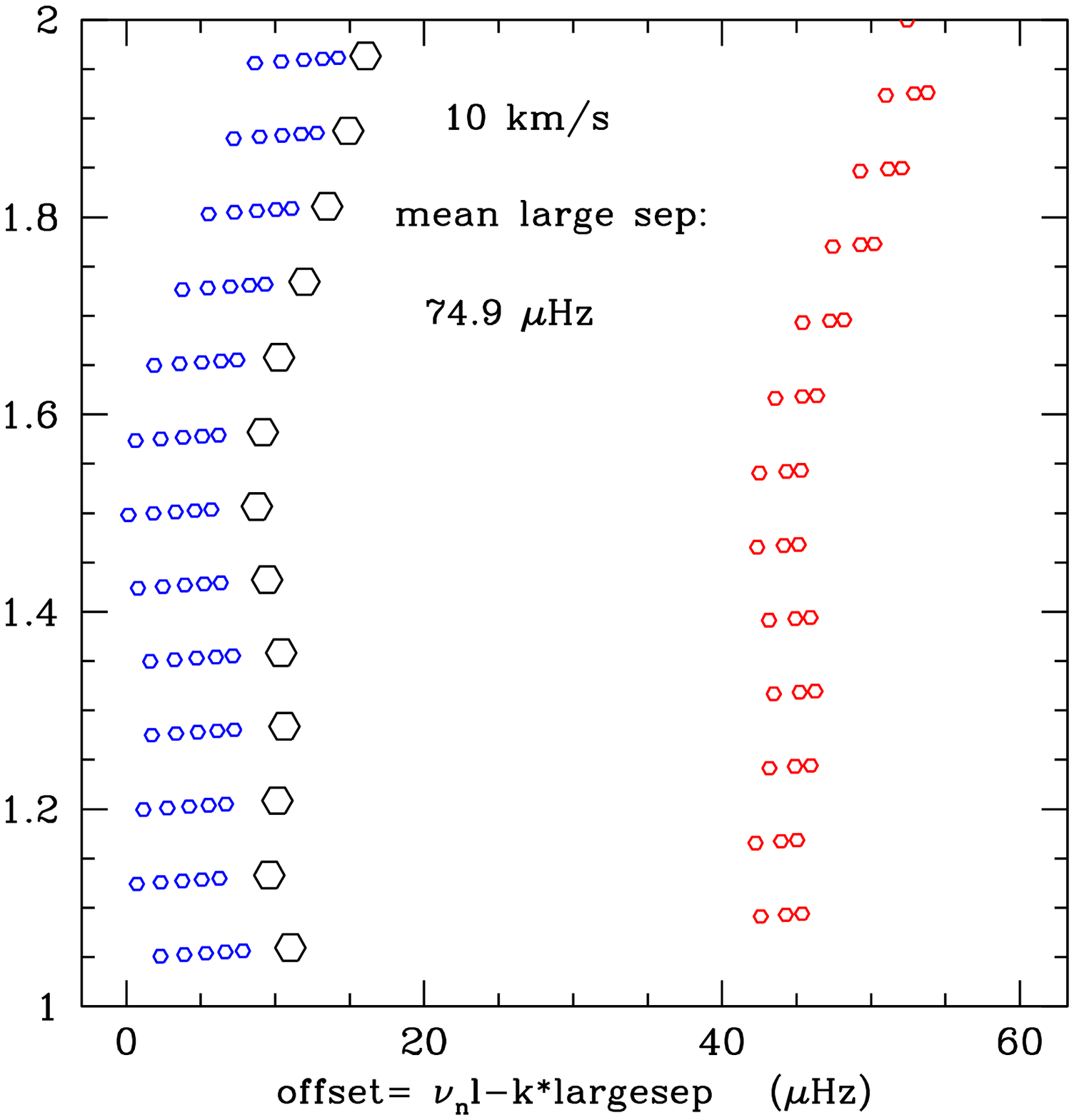}}
\resizebox*{0.5\hsize}{!}{\includegraphics*{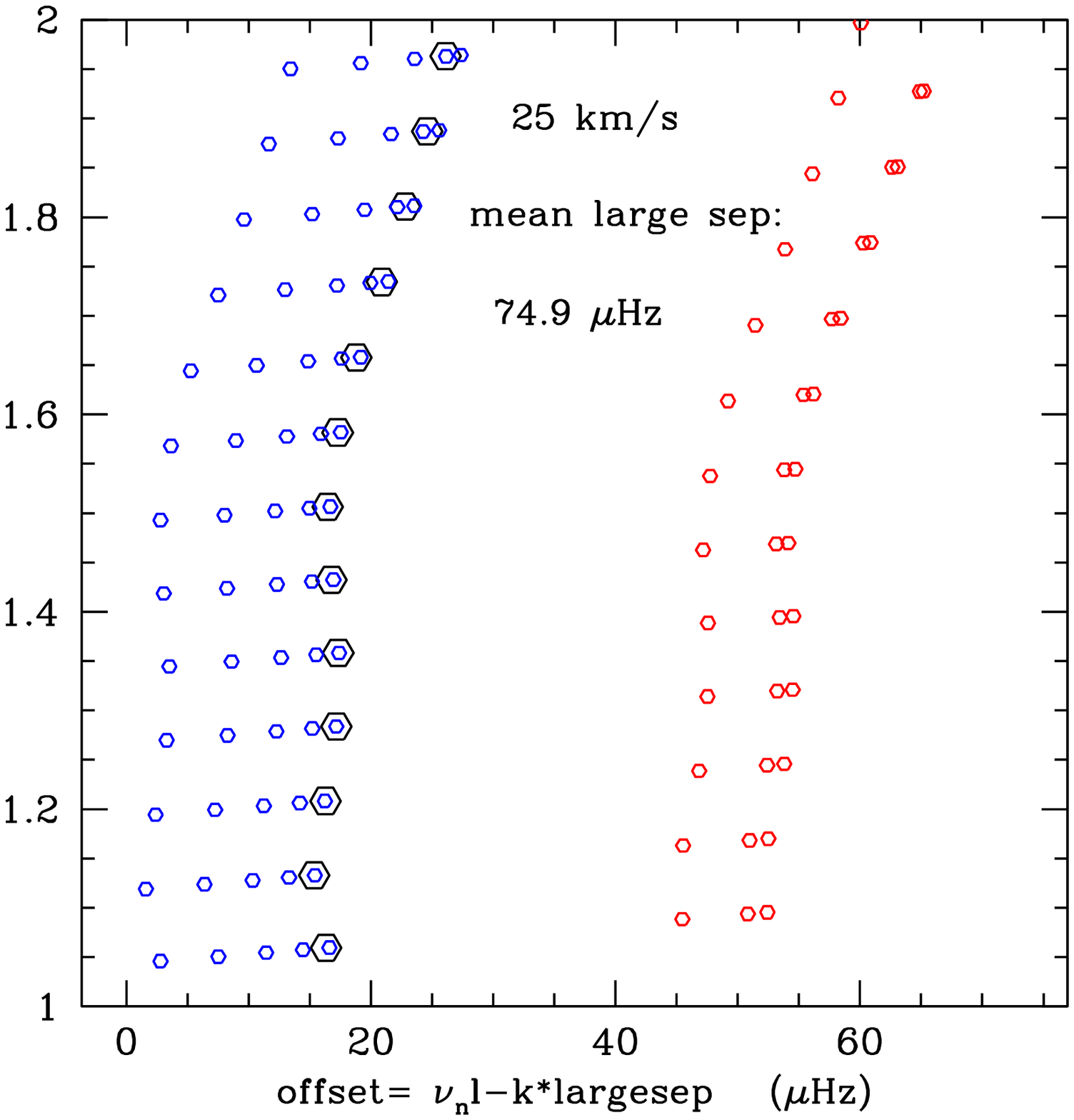}}
\caption{\small   Echelle diagrams for a model of HD181420:   rotational velocities
$v= 10$ km/s  {\bf (left)} and  25 km/s {\bf (right)}  are assumed   when computing the frequencies.
 From left to right in
each pannel,  ridges for $l=2,0, 1$ respectively appear. 
  }
\label{HD181}       
\end{figure} 

{\it Absolute frequencies   and echelle diagram} 	
The  echelle diagram displayed  in Fig.\ref{HD49} 
shows a  systematic shift between the observed and  theoretical 
$l=1$ ridges roughly independent of the frequency when the mean separation 
is taken to be the same in building the observed and theoretical echelle diagrams. 
This  discrepancy comes from the fact that
absolute values of the frequencies have not been included in the optimisation procedure to find an
optimal model. Increasing slightly the mean separation for building the echelle diagram 
for the model enables to 
perfectly match the low frequency part of the observed ridge as seen 
in  Fig.\ref{HD49}. The deviation of the theoretical ridge with the observed ones  remains  
 at high frequencies and might reflect  surface effects
  that are not  properly taken into account  in the numerical frequencies.

\subsubsection{HD181420: a fast rotator}
Analyses of the light curve of this star has been performed by  \cite{Barban09} and  
\cite{Gaulme09} who found 
two possible scenarii for  the mode identification. 
Results based on  a Bayesian approach  favors 
scenario 1 Benomar (2010, priv.com). This  is also the conclusion of  \cite{Bedding10} using  scaling properties. 
Focusing then on  scenario 1, one reproduces the  large separation and the small spacing $d_{01}$ 
with  a $1.36 M_\odot$  stellar model  assuming a core overshoot of $0.2 Hp$
 as well as with a  $1.37 M_\odot$  stellar model assuming no overshoot,
everything else being the same in particular  microscopic 
diffusion  and  rotation  are not included \cite{goupil09b}.
A secondary oscillation component is seen  in the observed 
large separation that is not reproduced by any models \cite{goupil09b}, Michel (2010 priv.com).   
The  'period' of this oscillation corresponds to the base of the 
convective zone  of the above  models,  but   
 the   reality of this secondary component is  not confirmed 
 \cite{benomar10phd}, Mosser (2010, priv. comm).
This star is a  'rapid' rotator compared to the Sun. Indeed with   a radius 
$R=1.66 R_\odot$  and  the observed mean rotational splitting  $\nu_{split} =  (3.\pm 1 )  \mu Hz$ \cite{Barban09},
one obtains a rotational velocity   $v  = 21.9 ± 7.3$ km/s  
that corresponds to a ratio of the centrifugal to the
gravitational accelerations of  $\epsilon =\Omega^2/(GM/R^3)  = 320 \epsilon_\odot$       !
Perturbation methods to compute the effect of rotation on the frequencies  nevertheless 
remain  valid 
for this rotation rate \cite{Suarez10}. 
The non-spherically centrifugal distortion causes asymetries of split multiplets that
 are  seen in   echelle diagrams already   for $v=10$ km/s at high frequency 
(Fig.\ref{HD181}, frequencies including rotating effects have been performed 
with  the WarM oscillation code).
Assuming  a uniform rotation velocity  of 25 km/s,
 the  $m=2$ components of the $l=2$ modes coincide with the $l=0$ mode for the lowest
 frequencies  whereas at high frequencies both $m=1$ and $m=2$ components are 
 mixed with the $l=0$ frequency.
Asymetries increase with frequency and are therefore larger at high frequencies where they 
 contribute to surface effects!
The difference between  the rotation frequency  measured at low
 frequency in a power spectrum  \cite{mosser10}
and the mean splitting  \cite{Barban09}
 is found ot be compatible with a latitudinal dependence of the surface rotation
  of the star (Ouazzani, 2010 priv. com).

\subsubsection{ HD49385: an evolved  star}
The light curve of this  star has been analysed by 
\cite{Deheuvels10c}  and \cite{Deheuvels10b}. \cite{Deheuvels10b} clearly put in evidence the existence of 
  an $l=1$ avoided crossing and showed that it produces a 
  characteristic  distortion of the $l=1$ ridge in the neighbour of this mode in an echelle diagram. 
A  clear explanantion of the  deformation  of the ridge  by the presence of an  avoided crossing 
 for a $l=1$ mode  has been provided by  \cite{Deheuvels10c}. The distorsion is caused by 
  the fact that the modes  propagate as p mode  in a surface  cavity and  as g mode in a central cavity. 
  For each   mode,  both cavities are  separated by an
  evanescent region  which acts as  a  coupling between the two regions. 
  The   magnitude of the ridge deformation is related to the 
  properties of the evanescent region.
 The   deformation of the ridge is then fitted  to constrain the stellar model. 
A model satisfying all  the constraints  simultaneously  then 
is difficult to obtain.   Only a change in
the chemical mixture from AGS05 to AGS09 is able to affect 
the evanescent region so as to provide a correct
ridge deformation \cite{deheuvels10phd}.


\section{ Kepler data and ensemble seismic investigations }

\begin{figure}[t]
\centering
\resizebox*{0.48\hsize}{!}{\includegraphics*{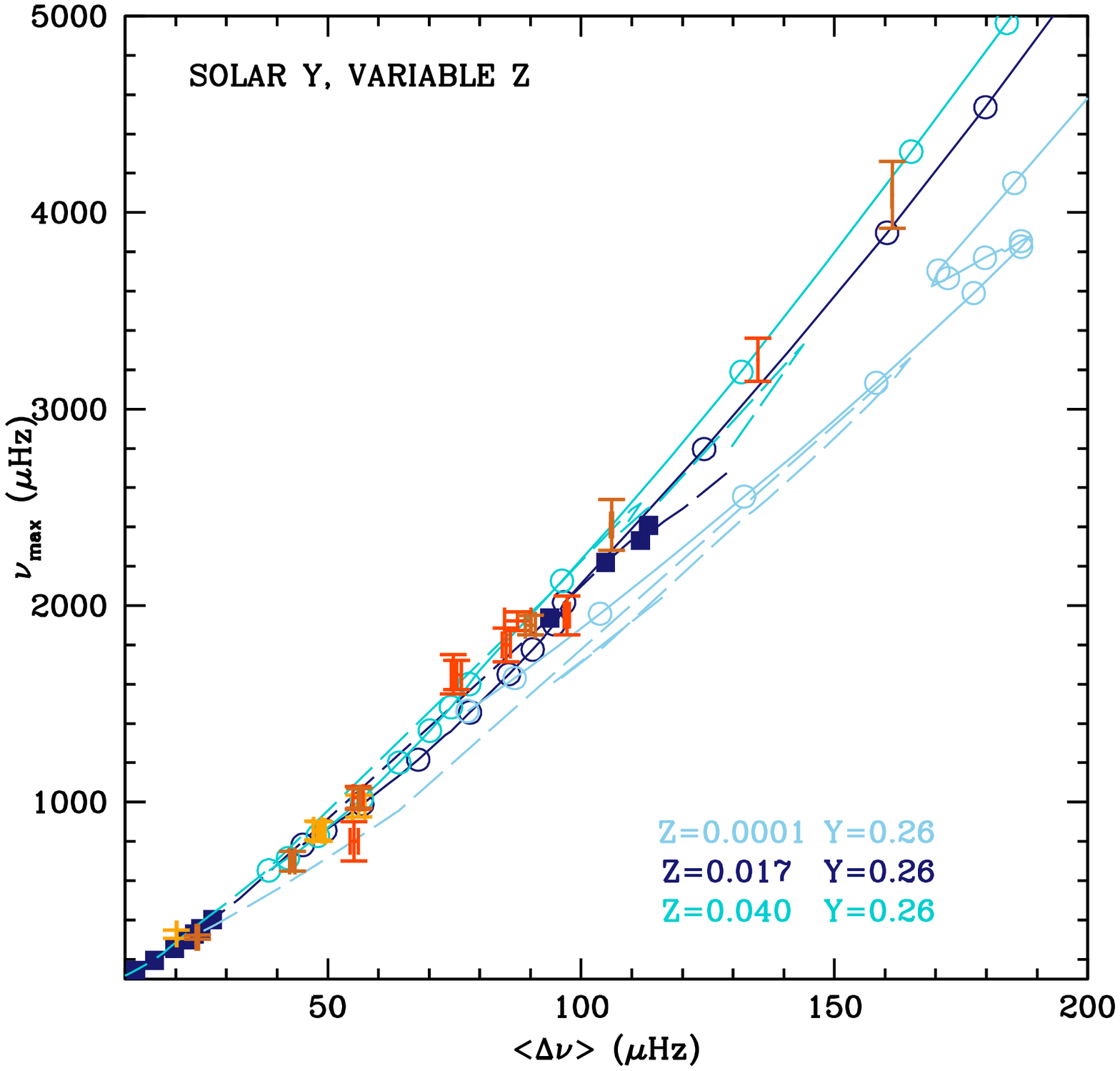}}
\resizebox*{0.48\hsize}{!}{\includegraphics*{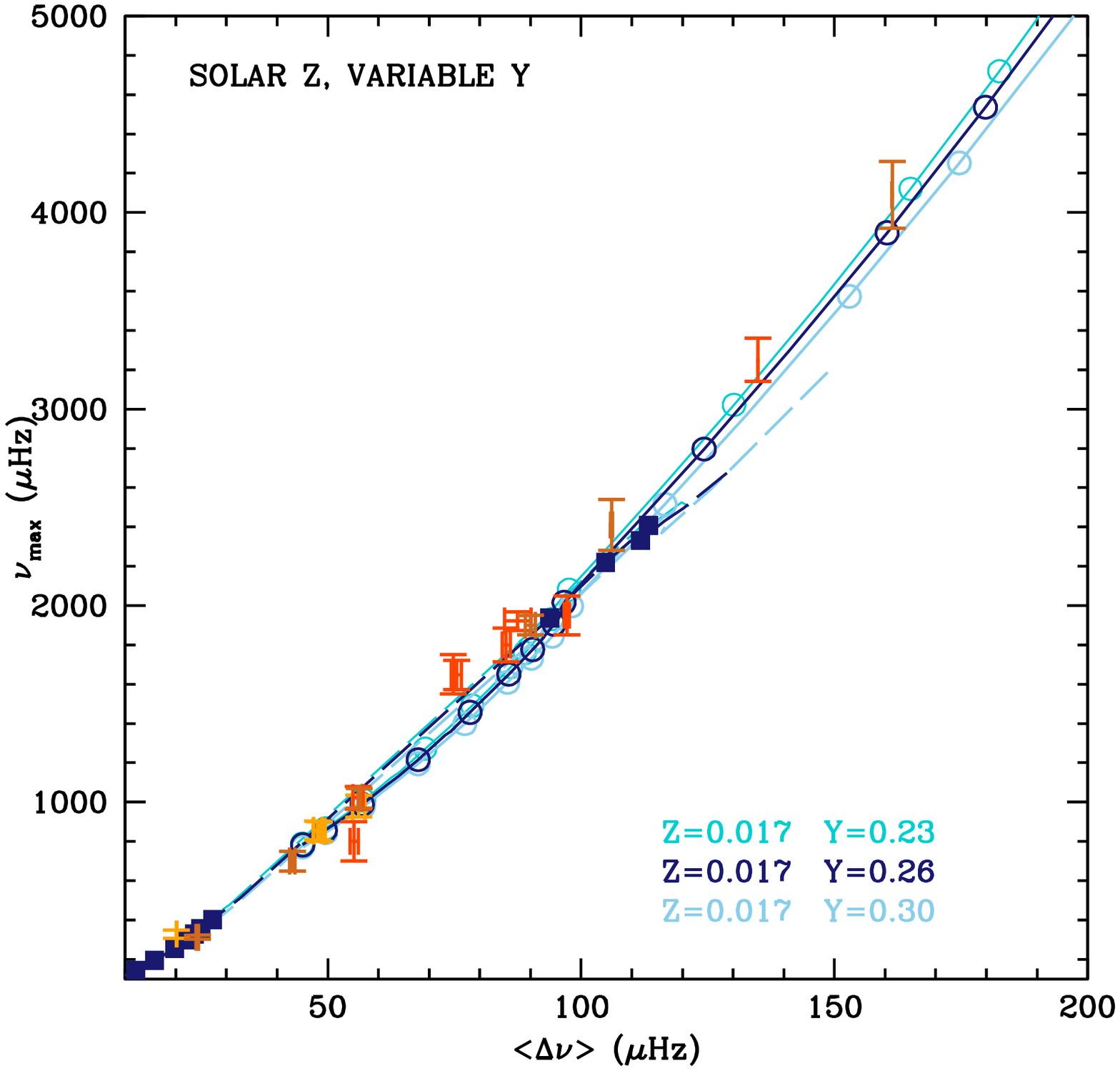}}
\caption{\small Frequency $\nu_{max}$ in function of the mean large separation $<\Delta \nu>$.
 Models (solid curves) have been computed with cesam2k code. Scaling laws are computed according to
  \cite{kjeldsen95} and  \cite{bedding03}.
   Data for  a few ground based observed stars    and some  CoRoT  targets  plotted with their errors bars 
represented by crosses.   }
\label{scal}       
\end{figure} 

The seismic part  of the space mission Kepler produces seismic data for a huge number of stars 
for which stellar parameters are in general not well known. 
Ensemble investigations   to derive stellar masses and radii 
then  rely on scaling seismic properties
 \cite{chaplin10}.
Fig.\ref{scal} shows   $\nu_{max}$ (frequency at maximum oscillation power in a power spectrum) in function of the  mean large separation 
 $<\Delta \nu>$ for   series 
 of stellar models (open circles) 
 assuming different chemical compositions  along evolutionary
 tracks from ZAMS (top  right corner)  to TAMS (down left corner). 
 Data for  a few ground based
observed stars are overplotted as well as  CoRoT  targets.  
Due to NASA data policy,  Kepler  data are not shown.  
large 
 The error bars indicate that one cannot distinguish between  metallicities 
  $ Z=0.017$ and $Z=0.04$ (with $Y=0.26$)  for instance, 
nor between  $Y=0.23$ and $Y=0.30$ ($Z=0.017$). This leads to small
uncertainties on  masses and radii derived from these scalings.
This is a crucial issue for   the ESA project PLATO 
 \cite{Catala10} that  aims at detecting  and studying earth-type exoplanets 
 and therefore  requires the determination of stellar masses  of exoplanet host stars  with an
accuracy of about 10-15$\%$. This means that it will be mandatory to  determine  the metallicity 
and helium abundances  $Z/X$ and $Y$  by other means; for instance $Z/X$ by
spectroscopy and $Y$ by the properties of the frequency dependence of the large separation.
Another important  issue is  to establish the  physical origin of the $\nu_{max}$ scaling 
which sofar has only been conjectured \cite{Brown91} but has been validated with observations 
\cite{bedding03}.
A better understanding of this origin could  provide 
  the explicit dependence of the scaling relations on metallicity and chemical
composition, rotation etc ...

\section{ Conclusion:}

   With the wealth of data from CoRoT and Kepler, we are  
 confronted  with seismologial studies of   a rich variety of low, and intermediate mass, main sequence 
  stars that   differ from  the Sun in their stellar parameters and internal
 structure. 
Although this will enrich considerably our understanding
of stellar physics and evolution,  this also generates  several 
difficulties not encountered with the Sun. Some were expected such as  
 inaccurate determination of stellar parameters, degeneracy in parameter space and 
 resulting non unicity of stellar models
but some other problems were not really   
 expected  such as ambiguities in the mode identification due to 
broad linewidths  and fast rotation (compared to the Sun) for F stars.  So
as far as  probing internal structure by means of seismology of solar-like stars is concerned, 
  we are therefore still at the beginning of the learning phase.  
  Significant advances in stellar physics and solid conclusions about the open issues discussed in the present
paper will require  homogeneous  detailed seismic studies for  a larger number of individual 
 stars that has been done so far.
 New roads also develope such as ensemble studies
   and scaling procedures and    additional observational seismic constraints
   start to exist such as  accurate 
 mode amplitudes   and linewitdhs  and their variations with frequency 
that were  not available from ground. 
Eventually  studying   a star  and its planets  as a global system  is certainely 
the issue that  must adressed in the future. In that framework,
 the perspective  offered by the ESA project 
PLATO  which  has precisely this aim     is a strong  motivation   for the seismic community  to pursue its
efforts   and  in turn PLATO will grandly benefit from all the forthcoming advances in the field.

\ack { We gratefully thank  our colleagues 
 K. Belkacem,  J. Ballot, B. Mosser,  C. Barban, 
 T. Corbard, D. Reese, O. Creevey, A. Baglin, E. Michel, T. Appourchaux, C. Catala, A. Mazumdar for providing useful information and 
 fruitful discussions when preparing this review. We also thank 
  P. Morel and J. Christensen-Dalsgaard for providing public
 evolutionary and oscillation codes respectively  that were used in the present review. We ackowledge financial
 support from CNES and the  ANR SIROCO}

\section{References}

\bibliographystyle{iopart-num}
\bibliography{aix_v12b}

\end{document}